# Ballistic spin resonance

S.M. Frolov[1], S. Lüscher[1], W. Yu[1], Y. Ren[1], J.A. Folk[1], W. Wegscheider[2]

[1]*Department of Physics and Astronomy, University of British Columbia, Vancouver, BC V6T 1Z4, Canada*

[2]*Institut für Angewandte und Experimentelle Physik, Universität Regensburg, Regensburg, Germany*

**The phenomenon of spin resonance has had far reaching influence since its discovery nearly 70 years ago.[1] Electron spin resonance (ESR) driven by high frequency magnetic fields has informed our understanding of quantum mechanics, and finds application in fields as diverse as medicine and quantum information.[2] Spin resonance induced by high frequency electric fields, known as electric dipole spin resonance (EDSR), has also been demonstrated recently.[3,4,5,6,7,8] EDSR is mediated by spin-orbit interaction (SOI), which couples the spin degree of freedom and the momentum vector. Here, we report the observation of a novel spin resonance due to SOI that does not require external driving fields. Ballistic spin resonance (BSR) is driven by an internal spin-orbit field that acts upon electrons bouncing at gigaHertz frequencies in narrow channels of ultra-clean two-dimensional electron gas (2DEG). BSR is manifested in electrical measurements of pure spin currents[9] as a strong suppression of spin relaxation length when the motion of electrons is in resonance with spin precession. These findings point the way to gate-tunable coherent spin rotations in ballistic nanostructures without external a.c. fields.**

Harnessing spin-orbit interaction is a promising route to achieving spin manipulation in a spintronic circuit.[10] SOI makes electron trajectories spin-dependent, which can lead to spatial spin separation and spin accumulation when the interaction is strong.[11,12,13,14]

Conversely, an electron's spin is influenced by its trajectory through an effective magnetic field $\vec{B}^{so}(\vec{k})$ due to SOI that depends on the momentum vector $\vec{k}$.[15] A trajectory $\vec{k}(t)$ that is periodic in time gives rise to an oscillating field $\vec{B}^{so}(t)$. Recent experiments have shown that spin rotations driven by such periodic motion can result from high frequency electric fields (EDSR).[3,4,5,6,7,8]

An alternative mechanism for generating a periodic trajectory, and consequently an oscillating spin-orbit field, is specular scattering of an electron between two parallel boundaries in a conducting channel (Fig.1c, inset). A spin resonance at frequency $f_0$ can be expected from this ballistic motion when the frequency of a typical bouncing trajectory matches the spin precession frequency in a magnetic field $B_0$:

$$f_0 \equiv \frac{g\mu B_0}{h} \sim \frac{v_F}{2w}, \quad (1)$$

where w is the width of the channel and $v_F$ is the Fermi velocity determined by the density of 2D electrons $n_s$: $v_F = \hbar\sqrt{2\pi n_s}/m^*$. The bouncing frequency can easily reach 10's to 100's of GHz in micron-scale ballistic channels of semiconductor 2DEG, a frequency range that is experimentally challenging to access in ESR or EDSR, and especially difficult on a chip.

We detect ballistic spin resonance by injecting electrons into a narrow channel of high mobility ($\mu$=4.44·10$^6$ cm$^2$/Vs at $n_s$=1.1·10$^{11}$ cm$^{-2}$ and T=1.5K) GaAs/AlGaAs 2DEG through a spin-selective quantum point contact (QPC) (Fig.1a,1b).[16,9,17] The charge current is drained on the left end of the channel. Diffusion of the accumulated spin polarization generates a pure spin current to the right of the injector, and a nonlocal voltage due to this spin current is measured by a detector QPC located 7-20μm down the channel.[18,19] Data presented here are from 3 channels of length 100μm defined using electrostatic gates. The mean free path in the channels, $\ell$, significantly exceeds the channel width ($\ell = 4 - 20 \mu m$ for w=1μm, and $\ell = 20 \mu m$ for w=3μm). The relaxation length of the spin current ranges from 10's of microns away from resonance to just a few microns on resonance.

Low frequency lock-in measurements in a dilution refrigerator were performed in magnetic fields, $B^{ext}$, up to 10 T in the plane of the 2DEG. The external magnetic field plays several roles. The first is to break the spin degeneracy of one-dimensional conductance subbands in the injector and detector QPC's, setting the direction and magnitude of QPC polarization, P. The second is to define the quantization axis and precession frequency for spins as they travel through the channel. The out-of-plane component, $B^{ext}_z$, could be controlled independently; this component modifies the spectrum of ballistic trajectories by bending them in cyclotron orbits.

Spin dynamics in a GaAs/AlGaAs 2DEG are governed by the total magnetic field, including both the external field and an effective field associated with spin-orbit interaction, $\vec{B}^{tot}(\vec{k}) = \vec{B}^{ext} + \vec{B}^{so}(\vec{k})$. The spin-orbit field is dominated by first-order contributions arising from bulk (Dresselhaus, β) and structural (Rashba, α) inversion asymmetry: [20,21]

$$\frac{1}{2}g\mu_B \vec{B}^{so}(\vec{k}) = (\alpha - \beta)k_y \hat{x} - (\alpha + \beta)k_x \hat{y}, \qquad (2)$$

where $\hat{x}$ and $\hat{y}$ define the 2DEG plane and correspond to [110] and [$\bar{1}$10] crystal axes respectively.

Qualitatively, Eq.2 shows that motion in the $\hat{y}$ direction induces an effective spin-orbit field in the $\hat{x}$-direction, $B^{so}_x$; motion in the $\hat{x}$ direction induces $B^{so}_y$. As shown in Fig.1, electron trajectories that connect the injector with the detector consist of rapid bouncing along $\hat{y}$ leading to a periodic $B^{so}_x$, while diffusive motion along $\hat{x}$ gives slower random changes in $B^{so}_y$. Spin resonance requires a periodic field transverse to the external field. The lack of periodicity in $B^{so}_y$ implies no resonance when $\vec{B}^{ext}$ is applied along $\hat{x}$, and indeed none was observed. Instead, the nonlocal signal increases monotonically with $B^{ext}_x$, reflecting primarily the increase in QPC polarization (Fig.1c).[9]

When $\vec{B}^{ext}$ is applied along $\hat{y}$, however, the periodic field $B^{so}_x$ leads to ballistic spin resonance and a collapse of the nonlocal signal at the field $B^{tot}=B_0$ of Eq.1 (Fig.1c). Similar to conventional cw-ESR, this resonance leads to a rapid randomization of spin direction, causing a strong suppression of the spin current and a collapse of the nonlocal voltage across the detector. The signal disappears completely in a 1 μm wide channel at $B^{ext}_y$=6-8T (Fig.1c), indicating that the spin current has completely relaxed before reaching the detector QPC 20μm away. The center of the resonance dip near $B^{ext}_y$=7T implies an electron density of $n_s$~0.8·$10^{11}$cm$^{-2}$ (Eq.1) that is close to the bulk value, and consistent with Hall measurements in the channel. Direct measurements of the spin relaxation length confirm that the dramatic suppression of the nonlocal signal near 7T is due to spin resonance (see Supplementary Information).

Resonant spin dynamics in ballistic channels are influenced by the details of electron trajectories; varying the parameters of typical trajectories changes the resonance conditions. For example, a lower electron density (i.e. Fermi velocity) leads to a lower resonant frequency $f_0$ (Eq.1), so the dip appears at a lower field $B_0$. As the density is lowered in a 1 μm wide channel (Fig.2a), $B_0$ shifts from 7T down to 4T. The magnetic field dependence of the spin signal is closely matched by Monte Carlo simulations of spin dynamics in the channel (Fig.2b, see Supplementary Information for simulation details).[22]

Wider channels also yield lower values of $f_0$. The BSR dip in a 3 μm wide channel (Fig.3) is expected at $B_0$=2.6T (Eq.1), a field too low to clearly resolve this resonance. At such low field, the primary visible effect of BSR is to counteract the increase in QPC polarization with applied magnetic field, giving a flatter spin signal in the range $B^{ext}_y$=0-5T compared to the 1μm wide channel (Fig.1c, Fig.3a). However, another dip can also be observed in the wider channel, at a field $B_0$=8T that corresponds to the third harmonic $3f_0$. Higher frequency components of the effective field are indeed expected when hard wall scattering within the channel leads to a square wave dependence for the velocity component $v_y(t) \propto B^{so}_x(t)$, with Fourier components at $(2N+1)f_0$ (Figs.3b,c,d). In principle this leads to a ladder of ballistic spin resonances extending to arbitrarily high external magnetic fields, but in practice the higher harmonics disappear quickly due to scattering. A clear third harmonic resonance is brought out in the simulation if one assumes a field independent QPC polarization P=1 and long mean free path $\ell$=30μm (Fig.3a Sim A). It is remarkable that trajectories even with such a long mean free path look so disordered to the eye (Fig.3b), but the resilience of the resonance to small amounts of disorder can be understood qualitatively from the square-wave character of the velocity (Fig.3c). More realistic simulation parameters, with shorter mean free path and finite perpendicular field, closely match the measured resonances (Fig.3a Sim B).

A small component of magnetic field applied perpendicular to the 2DEG, $B^{ext}_z \ll B^{ext}_{x(y)}$, bends electron trajectories in partial cyclotron orbits (Fig.4a). Such trajectories contain oscillating components along both $\hat{x}$ and $\hat{y}$, that give rise to periodic spin-orbit fields $B^{so}_y(t)$ as well as $B^{so}_x(t)$. A periodic $B^{so}_y(t)$ causes ballistic spin resonance when the large external field is applied along $\hat{x}$ (Fig.4c), in contrast to the case of small perpendicular field (Fig.1c) when BSR was only observed in $B^{ext}_y$. When the perpendicular field makes the cyclotron radius, $r_c$, less than the channel width, $r_c \equiv m^* v_F / eB^{ext}_z < w$, the characteristic bouncing frequency increases as electrons follow skipping orbits that don't extend across the entire channel (Fig.4a). This leads to an increase in the resonance field $B_0$ observed in the measurement and matched by simulation (Figs.4b,4e).

The qualitative effects of $B^{ext}_z$ can be seen in Fourier analyses of electron momenta along $\hat{x}$ and $\hat{y}$ for realistic trajectories (Figs.4d,4e). The spectrum of $v_x(t)$ shows no finite-frequency peak at $B^{ext}_z=0$, but a spectral component begins to emerge at 45GHz above $B^{ext}_z \sim 10$mT. This explains the appearance of a $B^{ext}_x$ spin resonance with increasing perpendicular field in Fig.4c. Above $B^{ext}_z \sim 40$mT the cyclotron radius falls below the channel width, so most trajectories to not cross the channel and the resonant frequency approaches the cyclotron frequency as the field is increased. Higher harmonics are quickly suppressed due to a transition from switching (square wave) to smooth changes in $B^{so}_x(t)$.

Equation 2 shows that a partial cancellation of bulk and structural asymmetry terms in the spin-orbit interaction may lead to different values of $B^{so}_x$ and $B^{so}_y$. Figures 2 and 3 show BSR for $B^{ext}_y$, which depends primarily on $B^{so}_x \propto (\alpha-\beta)k_F$. The resonance in $B^{ext}_x$, in Fig.4c, depends primarily on $B^{so}_y \propto (\alpha+\beta)k_F$. A comparison of these data sets with simulation enables both $\alpha$ and $\beta$ to be estimated: the best match is found with $\alpha = 1.3 \cdot 10^{-13}$eV·m and $\beta = -0.7 \cdot 10^{-13}$eV·m, implying a moderate anisotropy in the spin-orbit field. These values have been used in the simulations throughout the paper.

Results presented here demonstrate that the effect of spin-orbit interaction on spin coherence can be dramatically amplified with a geometric resonance. In quasi-diffusive structures like the channels shown here, spin polarization is randomized at the resonance. Coherent spin rotations could be achieved by the same mechanism if the effects of scattering were small and the number of ballistic cycles were well-defined, as is the case in the transverse electron focusing geometry.[16,12]

Acknowledgements. We thank M. Berciu, M. Duckheim, J.C. Egues, D. Loss, and G. Usaj for useful conversations. Work at UBC supported by NSERC, CFI, and CIFAR. W.W. acknowledges financial support by the Deutsche Forschungsgemeinschaft (DFG) in the framework of the program "Halbleiter-Spintronik" (SPP 1285).

Correspondence and requests for materials should be addressed to J.A.F. (jfolk@physics.ubc.ca).

**Fig.1. Pure spin current measurements reveal the ballistic spin resonance.**
**a**, schematic of the pure spin current measurement geometry. 2D electron gas (light background) is depleted under the gates (dark grey) to define the injector and detector QPC's and the diffusion channel. The gate pattern in the schematic is simplified for clarity. **b**, optical image of a measured device. Brighter colour indicates CrAu gates. Gate voltages $V_g^{inj}$ and $V_g^{det}$ are used to tune QPC's to the spin-polarized plateaus in high magnetic field; $V_g^\Lambda$ adjusts the length of the channel. Scanning electron micrograph of a prototype device details the injector area. **c**, Magnetic field dependence of the nonlocal voltage $V_{nl}$ measured at $G_{inj}=G_{det}\sim 1e^2/h$ for field $B^{ext}$ applied along $\hat{x}$ and $\hat{y}$. The channel width w=1μm, T=300 mK, $V_{ac}$=50μV, injector-detector spacing $x_{id}$=20μm. Inset: the effective spin-orbit field $B^{so}$ oscillates along $\hat{x}$ but is constant along $\hat{y}$ for an ideal specular trajectory.

Fig.2. **Tuning the ballistic spin resonance with electron density**. **a**, Magnetic field $B^{ext}_y$ dependence of the nonlocal spin signal for three electron densities in a 1 μm wide channel. T=500mK, $V_{ac}$=50μV, $x_{id}$=6.7μm. Data for the initial density is multiplied by 3 for better comparison to other densities (much

lower channel resistance at this density led to a lower nonlocal voltage). **b**, Monte Carlo simulations of the spin signal, described in detail in the Supplementary Information. Simulations for three Fermi velocities are shown: $v_F$=1.26·10$^5$m/s, 1·10$^5$m/s and 7·10$^4$m/s. $v_F$=1.26·10$^5$m/s was fixed by Hall measurements in the channel that gave density $n_s$=8.5·10$^{10}$cm$^{-2}$; the other two were chosen to match measured BSR dip locations, and correspond to densities $n_s$=5.5 and 2.5·10$^{10}$cm$^{-2}$ respectively.

Fig.3. **Third harmonic resonance in a wider ballistic channel. a**, Measured (circles) and simulated (lines) spin signal in a 3μm wide channel (T=500mK, $V_{ac}$=50μV, $x_{id}$=6.7μm). Nonlocal voltage is reduced compared to signals in 1μm wide channels due simply to wider channel (Eq.S1). Sim A: Simulation parameters tuned to show two dips clearly. $v_F$=1.1·10$^5$m/s, $\ell$=30μm, $B^{ext}_z$=0mT. In addition, injector and detector polarization were fixed at P=1 for this simulation to show the low-field dip. Sim B: realistic parameters approximate data: $v_F$=1.1·10$^5$m/s, $\ell$=10μm, $B^{ext}_z$=20mT. **b**, a simulated segment of an electron trajectory in a 3 μm wide channel (parameters as in Sim A above). **c**, the component of electron velocity $v_y$ for the trajectory in **b** is reminiscent of a square wave, changing sign each ~30 ps (the crossing time of the channel). **d**, Power spectrum of $v_y$ shows a comb of peaks that correspond to BSR at odd multiples of the primary frequency $f_0$≈16.7GHz.

Fig.4. **Out-of-plane magnetic field modifies resonance.** The experimental parameters for this figure are: w=1μm, T=500mK, $V_{ac}$=50μV, $x_{id}$=6.7μm. **a**, cartoons of trajectories for cyclotron radius greater than (upper) and less than (lower) channel width. Effective spin-orbit fields (orange arrows) oscillate along both $\hat{x}$ and $\hat{y}$. **b**, Resonance shifts to higher magnetic field $B_0$ when cyclotron orbits do not cross channel, shown here for $B^{ext} \parallel \hat{y}$. **c,** Resonance emerges

also for $B^{ext} \| \hat{x}$ when finite $B^{ext}_z$ bends trajectories, giving rise to periodicity in $B^{so}_y(t)$. Inset: Simulation for realistic parameters ($v_F$=1·10$^5$m/s, $\ell$=10μm). **d,e**, Power spectra for $v_x$ (**d**) and $v_y$ (**e**) with parameters from the inset of **c**. Dashed lines in **d**(**e**) show range of experimental data in **c**(**b**). Dotted line in (**e**) indicates cyclotron frequency. Color scale denotes logarithm of spectral power.

**Figure 1**

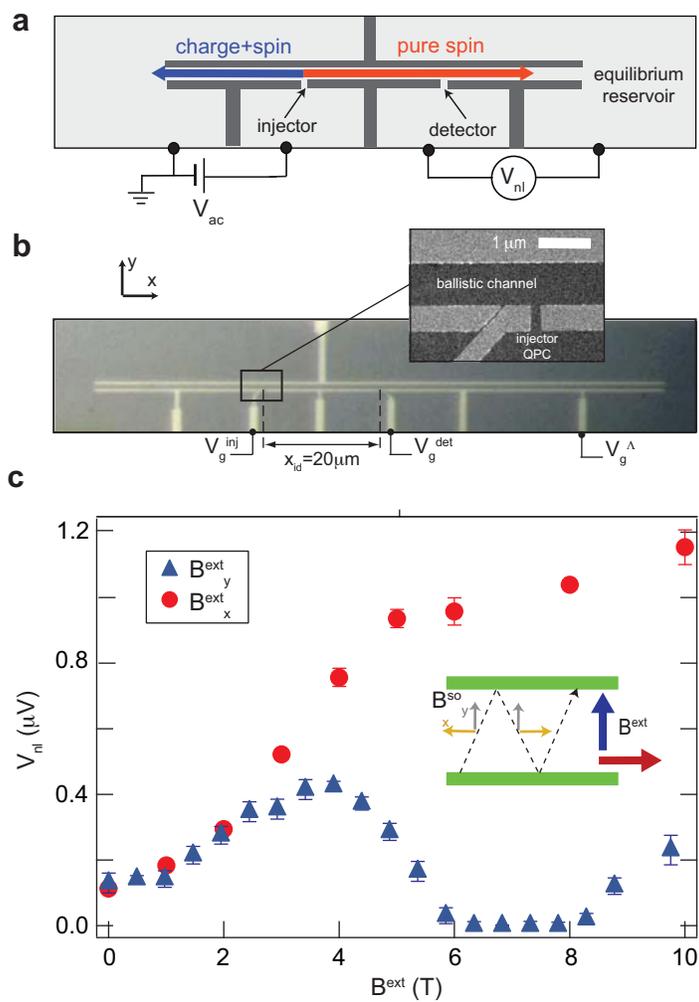

**Figure 2**

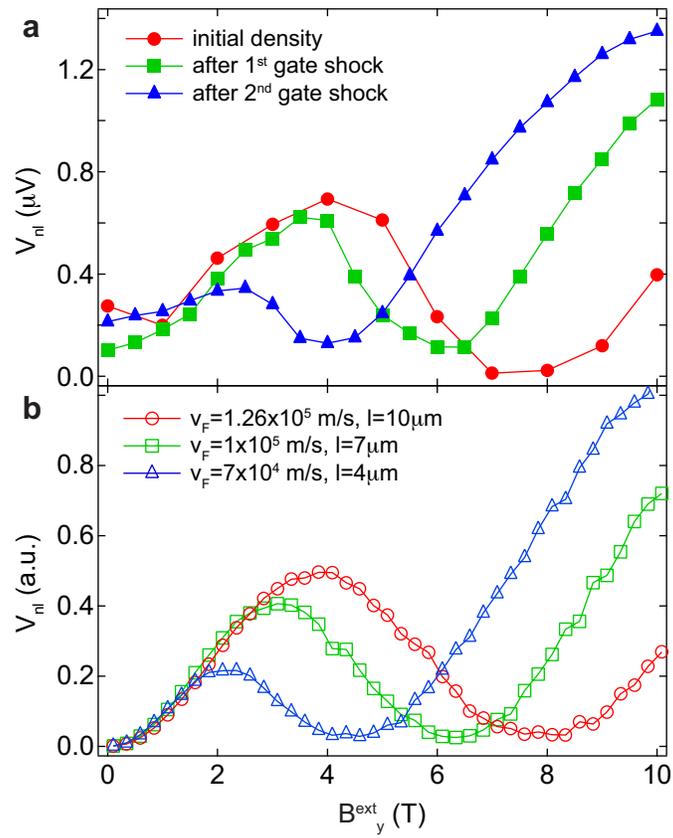

**Figure 3**

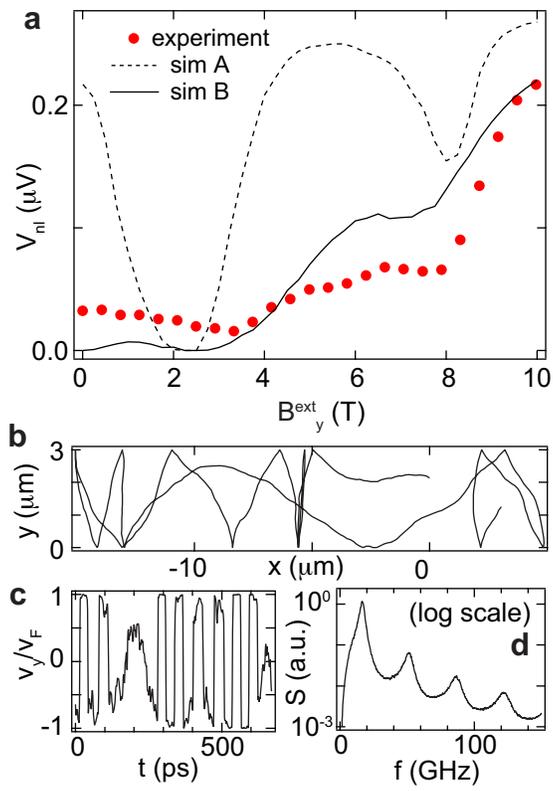

**Figure 4**

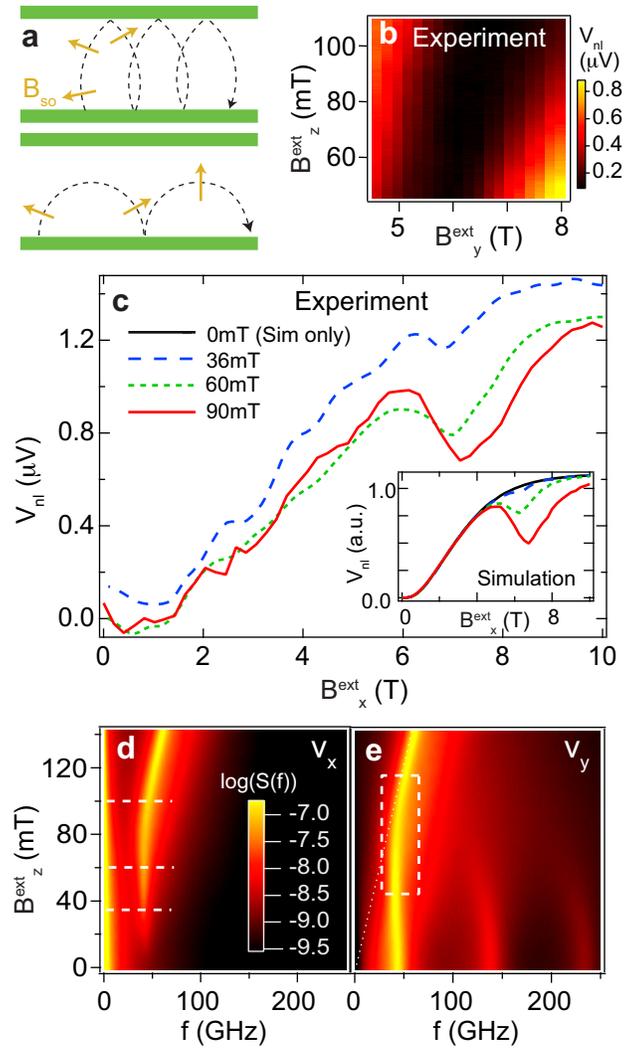

# Supplementary Information: Ballistic spin resonance

S.M. Frolov[1], S. Lüscher[1], W. Yu[1], Y. Ren[1], J.A. Folk[1] and W. Wegscheider[2]

[1]Department of Physics and Astronomy, University of British Columbia, Vancouver, BC V6T 1Z4, Canada

[2]Institut für Angewandte und Experimentelle Physik, Universität Regensburg, Regensburg, Germany

## *Spin Relaxation Length Measurements*

A direct measurement of the spin relaxation length confirms that the suppressed signal for $B^{ext}_y$=4-10T (compared to $B^{ext}_x$) in Fig.1c is due to enhanced spin relaxation. The method is summarized below.[S1] The nonlocal voltage reflects the chemical potential difference between the two spin populations at the detector, located a distance $x_{id}$ from the injector. For a given spin relaxation length, $\lambda_s$, the expected $V_{nl}(x_{id})$ due to spin can be calculated by solving the 1D diffusion equation for a finite channel:

$$V_{nl}(x_{id}) = K \frac{\lambda_s}{w} \rho I_{inj} P_{inj} P_{det} \sinh \frac{L_r - x_{id}}{\lambda_s}, \qquad (S1)$$

where $K=1/[\sinh[L_r/\lambda_s] \cdot (\coth[L_r/\lambda_s]+\coth[L_l/\lambda_s])]$ is defined by the boundary condition of zero polarization at the left and right ends of the channel (distances $L_l$ and $L_r$ from the injector).

The dependence of $V_{nl}$ on either $L_r$ or $x_{id}$ can, in principle, be used to extract the spin relaxation length from Eq. S1. Different detectors give different $x_{id}$, but without independent knowledge of QPC polarization it is difficult to extract relaxation length from such a measurement. Instead, the flexibility of gate-defined channels in our measurement allowed $L_r$ to be changed in-situ without affecting the injector or detector, enabling a measurement of $\lambda_s$ that was independent of QPC polarization. Undepleting the $\Lambda$-gate (Figs.S1a,S1b,1b), reduced $L_r$ from $L_{r\text{-long}}$=70µm to $L_{r\text{-short}}$=40µm and caused the nonlocal signal to drop (Figs.S1c,S1d). The ratio in the signal for long and short channels, $\eta = V_{nl}(x_{id},L_{r\text{-long}})/V_{nl}(x_{id},L_{r\text{-short}})$, depends on $\lambda_s$ but does not depend on $P_{inj}P_{det}$ (Eq. S1). For the device measured here, Eq. S1 gives $\eta$=1.7 for no relaxation, $\eta$=1.3 for $\lambda_s$ = 30µm, and $\eta\sim1$ for $\lambda_s \ll L_{r\text{-short}}$.

Spin relaxation length measurements complimentary to the spin signal measurements in Fig.1c are presented in Fig.S1e. The relaxation length extracted from the ratio $\eta$ was $\lambda_s$ = 50 ± 10 µm for $\hat{x}$-oriented spins, nearly independent of field from $B^{ext}_x$=2-10T. For fields applied along $\hat{y}$, on the other hand, $\lambda_s$ collapsed above $B^{ext}_y$=4T, then partially recovered for $B^{ext}_y$>8T. The absence of a measurable nonlocal signal for $B^{ext}_y$=6T-8T and $x_{id}$=20µm implies an upper bound $\lambda_s$< 5µm within this range of fields. These measurements confirm that the nonmonotonic $B^{ext}_y$ dependence of the spin signal is due to spin relaxation inside the channel.

## *Other Explanations*

Alternative mechanisms for the suppressed nonlocal spin signal at high magnetic field were considered. They included a drop in the QPC polarization and an increase in the channel resistivity.

A dramatic nonmonotonic change in QPC polarization as a function of in-plane field would have been surprising given the smooth transition in conductance traces from 0-10T for both $B^{ext}_x$ and $B^{ext}_y$ (Fig.S2a). The quantum point contact polarization for $B^{ext}_x$ was extracted from the nonlocal spin signal in Ref.S1. In that measurement, the polarization increased when the magnetic field was applied along $\hat{x}$. An accurate spin current measurement of QPC polarization at the resonance in $B_{ext}\|\hat{y}$ was not possible because of the vanishing signal. However, a clear conductance plateau at $G=1e^2/h$ was observed in all QPC's at $B^{ext}>5T$ for both magnetic field directions (Fig.S2a), suggesting that no anomalous suppression occurs in the QPC polarization at magnetic fields for which $g\mu B>k_B T$. Furthermore, the spin relaxation length measurement described above would have been unaffected by a change in QPC polarization.

Charge transport through the channel showed no anomalous features in the broad field range 0-10T, for either field orientation. The resistance increased monotonically with in-plane field as shown in Fig.S2b, presumably due to compression of the 2DEG wavefunction. Similarly, thermoelectric measurements of heat transport through the channel (primarily mediated by electrons below 1K) revealed no unexpected features in the range $B^{ext}_y=3-10T$ (Fig.S3). These two observations prove that the collapse of the nonlocal signal as in Fig.1c was not due to an unexpected increase in charge scattering in the narrow field range $B^{ext}_y=6-8T$.

## *Changing the Electron Density*

In this experiment neither back gate nor top gate covering the channel was available. Instead, the density was changed by applying large negative voltages ($\approx-1V$) to the channel gates for a short period to "shock" the 2DEG. This reduced the number of charged dopant sites in a wide area surrounding the gates, thereby irreversibly lowering the density until the sample was brought up to room temperature. The density of electrons in the channel could be directly measured in devices with QPC's across the channel from the injector and detector QPC's. Hall measurements in such channels yielded densities of $0.8$-$0.9 \cdot 10^{11} cm^{-2}$ in a channel with resistivity $\rho=20$ $\Omega$/square, and $0.3$-$0.4 \cdot 10^{11} cm^{-2}$ for $\rho=120$ $\Omega$/square, implying that the mean free path was reduced from ~20 μm at the higher density to ~4 μm at the lower density.

A Hall measurement was not possible in the device used to demonstrate the shift of the resonance dip (Fig.2a). However, the density for each of the three traces in this figure could be estimated from the different values of ρ for each measurement. These densities were consistent with the Fermi velocities extracted from fits of the data to Monte Carlo simulations, see main text. The shifts to lower electron density in more resistive channels also caused a shift of the pinch-off points of the injector and detector QPC's to more positive gate voltages (Fig.S4).

### *Monte Carlo Simulations*

Monte Carlo simulations of electron trajectories in channels with the geometries of this experiment, and including appropriate models of disorder, shed more light on the ballistic spin resonance in a realistic device. These simulations have been included in Figs.2-4, and are being published in more detail elsewhere (Lüscher, Frolov, Folk, in preparation). Each simulation averages over thousands of electron trajectories, generated randomly including small-angle scattering corresponding to a particular mean free path and incorporating a particular out-of-plane magnetic field. Scattering off the walls includes a deviation from specularity with angular spread 0.25 radians. Details of the scattering did not significantly affect the results of the simulations.

The electron spins were initialized at the beginning of each trajectory along the external magnetic field, and evolved in the total magnetic field, $\vec{B}^{tot}(\vec{k}) = \vec{B}^{ext} + \vec{B}^{so}(\vec{k})$. The effective spin-orbit field was calculated using Eq.2 in the main text, with α=1.3·10$^{-13}$eV·m and β=-0.7·10$^{-13}$eV·m selected to best match the data. Average polarization along the different trajectories was then calculated as a function of distance in the channel, and from this length dependence a relaxation length was extracted. This relaxation length was plugged in to Eq.S1 to extract the nonlocal voltage. To match the experiment, the simulated spin polarization is multiplied by magnetic field-dependent

QPC polarization (measured in Ref S1) with the exception of Sim A in Fig.3a, where full QPC polarization is used.

**Figure S1. Spin relaxation length measurement.** The length of the channel is changed by depleting (**a**) or undepleting (**b**) the $\Lambda$-gate. The chemical potential difference between spin-up and spin-down (gradient red) decays to zero at channel ends due to large 2DEG reservoirs. **c**,**d:** nonlocal signal measured with polarized detector (w=1$\mu$m, $x_{id}$=20$\mu$m, T=300mK, $V_{ac}$=50$\mu$V) decreases when the $\Lambda$-gate is undepleted, suggesting a long spin relaxation length off resonance (**c**), and a much shorter spin relaxation length in vicinity of resonance (**d**). **e**, Magnetic field dependence of the ratio $\eta$ for $B^{ext}$ along $\hat{x}$ and $\hat{y}$. Ratio could not be calculated from $B^{ext}_y$=6-8T because no signal was visible.

**Figure S2. a,** Spin-resolved plateau $G_{inj}$=1$e^2$/h in the injector QPC develops smoothly with $B^{ext}_y$, providing further evidence that collapse of nonlocal signal from $B^{ext}_y$=6-8T is not due to lack of QPC polarization. Similar dependence was obtained when $B^{ext}$ was applied along $\hat{x}$. **b,** Channel resistivity increases monotonically with $B^{ext}_y$, indicating that charge scattering shows no unusual features from $B^{ext}_y$=6-8T, where the nonlocal signal is suppressed to zero.

**Figure S3.** Injector-detector gate voltage scans of the first and the second lock-in harmonics of the nonlocal signal measured at a magnetic field $B^{ext}_y$ **a,** below the resonance field, **b,** on resonance **c,** above the resonance field. Bright squares in the first harmonic correspond to the spin signal at the polarized odd conductance plateaus in the injector and detector QPC's. On resonance (panel **b**), only a weak thermoelectric signal due to Peltier heating can be discerned in the first harmonic, and this signal appears at the conductance steps rather than polarized plateaus (see Ref.S1). Joule heating by the injector current gave rise to a nonlocal thermoelectric voltage $V_{nl} \sim I_{inj}^2 S_{det}$ at twice the lock-in frequency, where $S_{det}$ is the thermopower of the detector QPC. This signal appeared when the detector conductance was tuned to transitions between plateaus ($S_{det}$ is zero at the plateaus), and was not suppressed by the spin resonance. This demonstrates that other nonlocal signals can be observed even when the spin signal is suppressed due to BSR. In this figure w=1μm, $x_{id}$=6.7μm, T=500mK, $V_{ac}$=50μV.

**Figure S4.** The conductance traces of the injector QPC corresponding to the three magnetic field dependences in Fig.2a shift to more positive gate voltages when the density is decreased.

# Figure S1

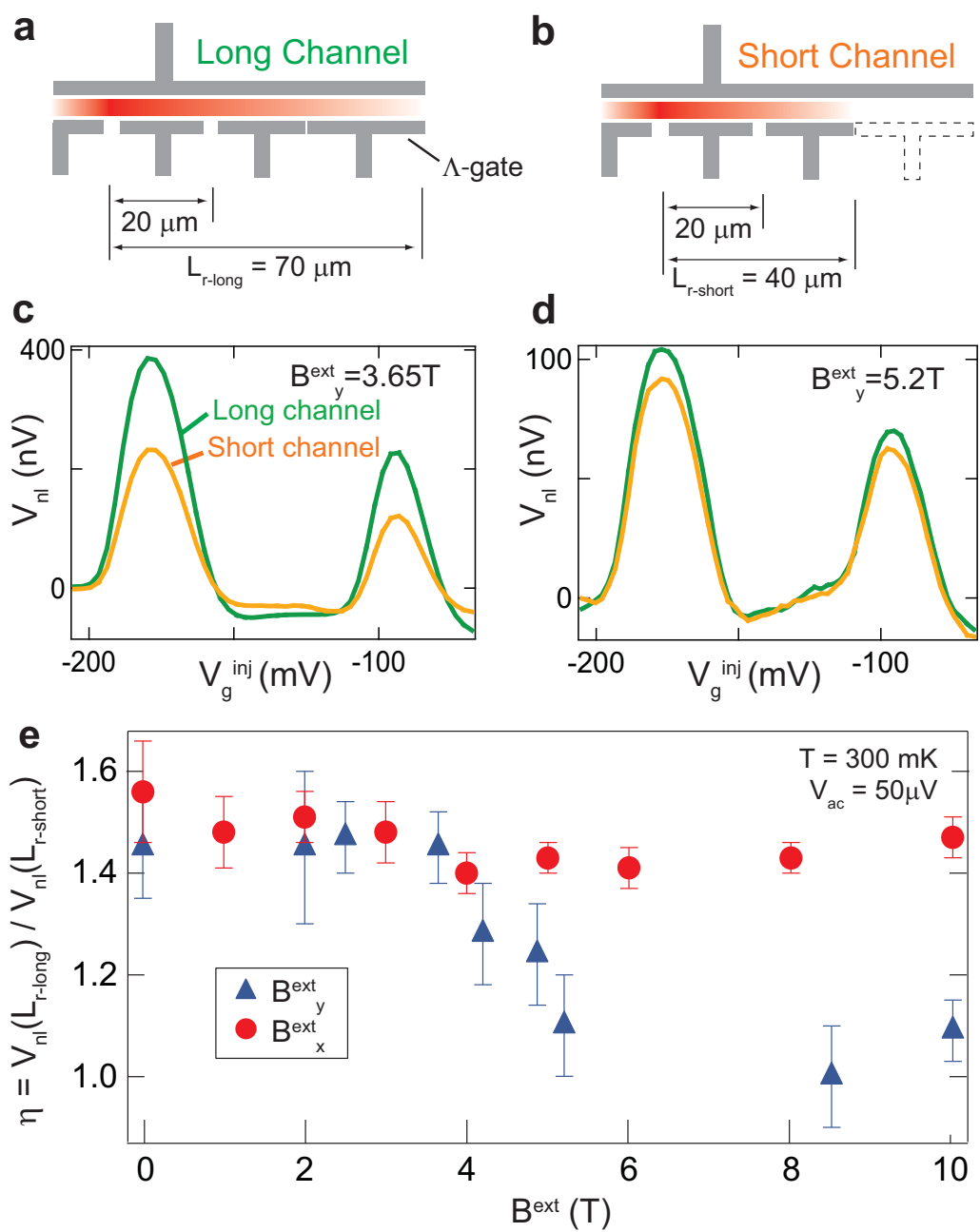

**Figure S2**

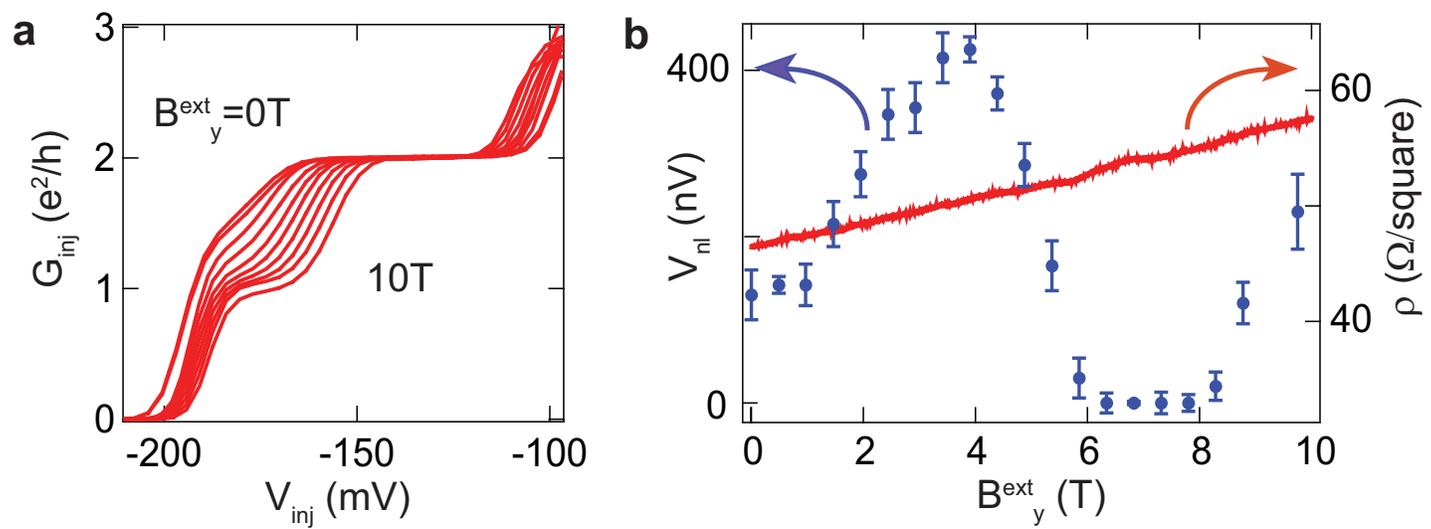

**Figure S3**

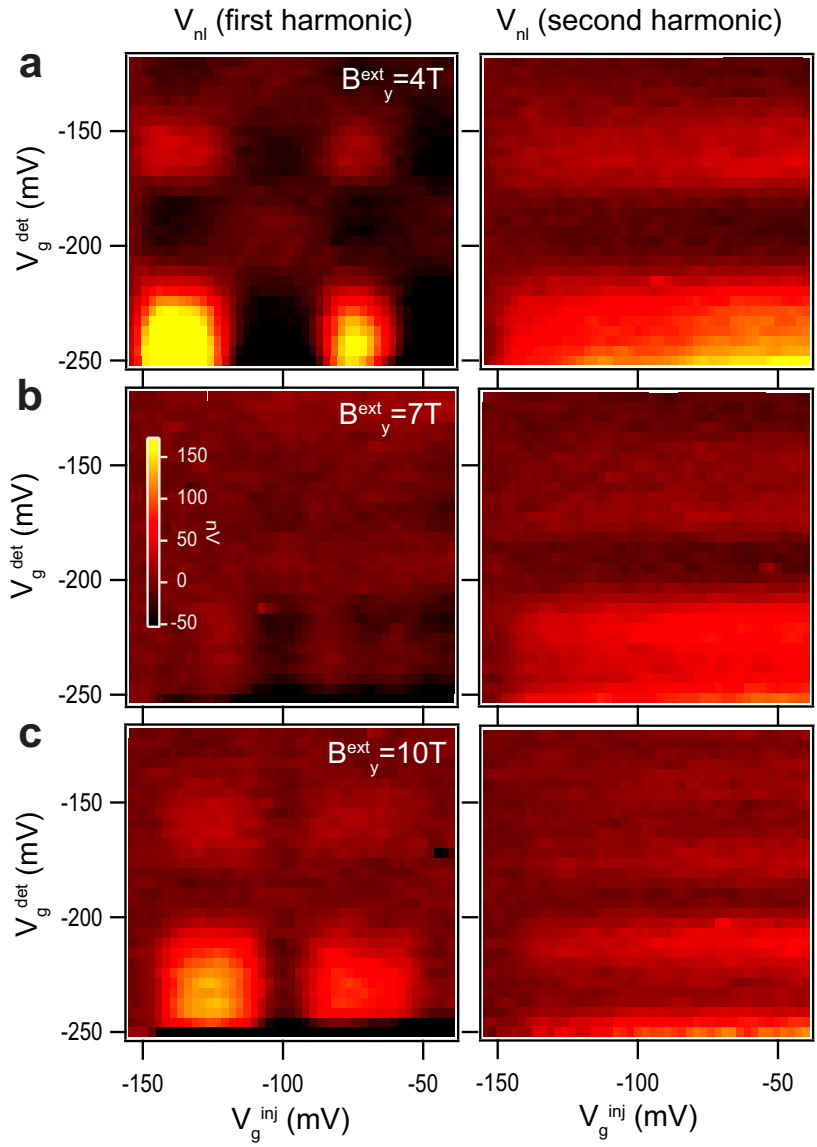

# Figure S4

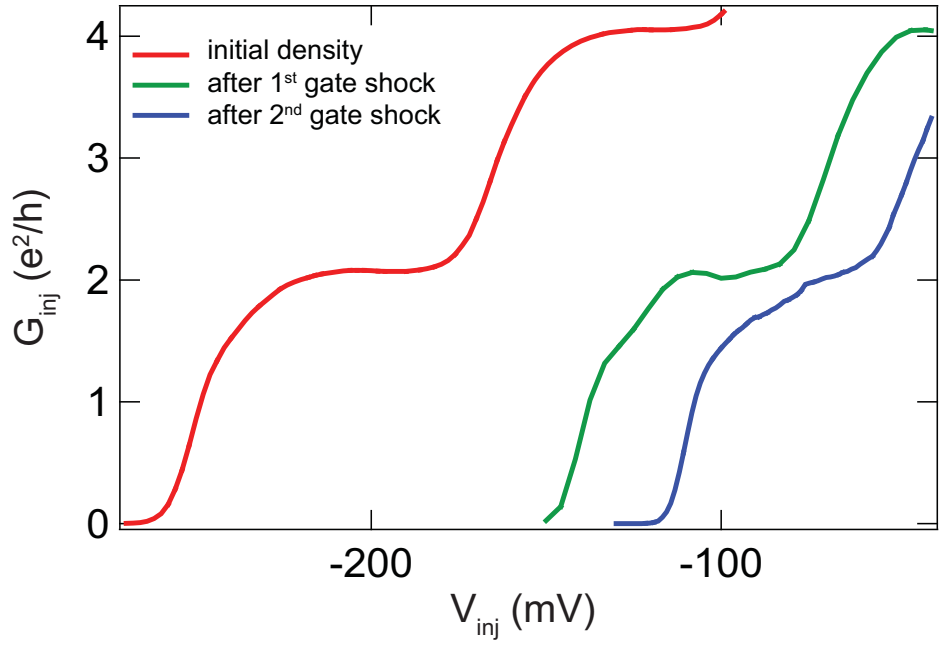